\documentclass[a4paper,fleqn,usenatbib]{mnras}

\usepackage{newtxtext,newtxmath}

\usepackage[T1]{fontenc}
\usepackage{ae,aecompl}

\usepackage{graphicx}	
\usepackage{amsmath}	
\usepackage{amssymb}	
\usepackage{nicefrac}   

\newcommand{\eg}{e.g. }

\newcommand{\rsh}{{r_{\rm sh}}}

\newcommand{\Msol}{\ensuremath{\rm{M}_{\odot}}}
\newcommand{\mns}{M_{\rm{NS}}}

\newcommand{\cms}{\rm{cm^2\ s^{-1}}}

\title[Impact of SASI on the pulsar spin]{Are pulsars spun up or down by SASI spiral modes?}

\author[R\'emi Kazeroni et al.]{
R\'emi Kazeroni,$^{1,2}$\thanks{E-mail: kazeroni@MPA-Garching.MPG.DE}
J\'er\^ome Guilet,$^{2}$
and Thierry Foglizzo$^{1}$
\\
$^{1}$Laboratoire AIM, CEA/DRF-CNRS-Universit\'e Paris Diderot, IRFU/D\'epartement d'Astrophysique, CEA-Saclay F-91191, France\\
$^{2}$Max-Planck-Institut f\"ur Astrophysik, Karl-Schwarzschild-Str. 1, D-85748 Garching, Germany 
}

\date{Accepted XXX. Received YYY; in original form ZZZ}

\pubyear{2016}

\begin{document}
\label{firstpage}
\pagerange{\pageref{firstpage}--\pageref{lastpage}}
\maketitle

\begin{abstract}
Pulsars may either be spun up or down by hydrodynamic instabilities during the supernova explosion of massive stars.
Besides rapidly-rotating cases related to bipolar explosions, stellar rotation may affect the explosion of massive stars in the more common situations where the centrifugal force is minor.
Using 2D simulations of a simplified setup in cylindrical geometry, we examine the impact of rotation on the Standing Accretion Shock Instability (SASI) and the corotation instability, also known as low-T/|W|.
The influence of rotation on the saturation amplitude of these instabilities depends on the specific angular momentum in the accretion flow
and the ratio of the shock to the neutron star radii.
The spiral mode of SASI becomes more vigorous with faster rotation only if this ratio is large enough.
A corotation instability develops at large rotation rates and impacts the dynamics more dramatically, leading to a strong one-armed spiral wave.
Non-axisymmetric instabilities are able to redistribute angular momentum radially and affect the pulsar spin at birth.
A systematic study of the relationship between the core rotation period of the progenitor and the initial pulsar spin is performed.
Stellar rotation rates for which pulsars are spun up or down by SASI are estimated.
Rapidly spinning progenitors are modestly spun down by spiral modes, less than $\sim 30\%$, when a corotation instability develops.
Given the observational constraints on pulsar spin periods at birth, this suggests that rapid rotation might not play a significant hydrodynamic role in most core-collapse supernovae.
\end{abstract}




\section{Introduction}
\label{sec:intro}

The study of the birth properties of pulsars is of considerable importance for the understanding of core-collapse supernovae. 
These properties provide meaningful information on the progenitors as well as the physics of the explosion.
In particular, the initial pulsar spin may be related to the core rotation period of the progenitor which can affect the dynamics of the core-collapse and the explosion.
A majority of young neutron stars are considered to be slow rotators, with periods of a few tens to a few hundreds milliseconds. 
Such a broad distribution can be inferred from observations of known pulsars whose age can be estimated \citep{vranesevic04,popov12} as well as population synthesis studies \citep{faucher06}.
This distribution sets a strong constraint on the explosion mechanism, suggesting that either explosions result from progenitors with slowly rotating cores at the point of collapse
or that there exists an efficient spin-down mechanism.

The inner angular momentum profile of massive stars is still poorly constrained. Recent asteroseismic observations of low-mass red giants \citep{beck12, mosser12} have revealed
that the core of these stars seem to rotate significantly slower than predicted by stellar evolution calculations \citep{cantiello14,deheuvels14}.
These computations including angular momentum transport due to magnetic torques predict pulsar spin periods of $10-15\,\rm{ms}$, rather fast compared to the observations \citep{heger05}.
The inclusion of internal gravity waves in these computations of stellar evolution could affect the distribution of angular momentum \citep{fuller15}.
Binary interaction during the progenitor lifetime may also influence the rotation profile \citep{sana12}.
Even in the last stage of stellar evolution, during core collapse, hydrodynamic instabilities can induce 
transverse mass motions which can impact the initial rotation period of the neutron star \citep{wongwathanarat10,wongwathanarat13}.

The most promising framework to explain core-collapse supernovae is the delayed neutrino-driven mechanism \citep{bethe85,janka16,mueller16}. In this scenario, a fraction of the copious number of neutrinos
emitted from the forming neutron star and the accretion layer interacts with the matter in the gain layer, where neutrino absorption dominates neutrino emission, to revive the stalled accretion shock.
Except at the low-mass end of supernova progenitors, massive stars do not explode in spherical symmetry \citep{liebendoerfer01,kitaura06}. Multi-dimensional simulations of
core-collapse supernovae have emphasized the supportive role of hydrodynamical instabilities for the shock revival. Two types of instabilities can break the spherical
symmetry of the collapse and lead to an asymmetric explosion: the neutrino-driven convection \citep{herant94, janka96} and the Standing Accretion Shock Instability
(SASI) \citep{blondin03}.

Simulating robust explosions from ab initio models remains a theoretical and numerical challenge.
Even though state-of-the-art 3D models seem to be close to the explosion threshold \citep{hanke13,melson15b,lentz15,janka16}, the explosion energies
obtained only cover the lower side of the range $10^{50}-10^{51}\,\rm{erg}$ inferred from observations \citep{mueller15b}. Several ingredients have been considered
for their ability to ease the explosion, such as rotation \citep{iwakami14b,nakamura14}, magnetic fields \citep{guilet11,obergaulinger14}, pre-collapse inhomogeneities \citep{couch13,mueller15}
and changes in neutrino-nucleon interaction rates \citep{melson15b}.
Our work focuses on the impact of stellar rotation and covers a wide range of rotation rates from slow to fast spinning cases, the former having received little attention in 3D models so far.
Rotation may facilitate standard core-collapse supernovae in several ways.
The critical neutrino luminosity threshold for explosion decreases with the specific angular momentum injected \citep{iwakami14b,nakamura14}. 
Rotation favours the development of prograde SASI spiral modes \citep{blondin07a,yamasaki08}, even if the centrifugal force is weak.
Above a certain rotation rate, the development of a strong one-armed spiral mode related to a corotation instability 
may help the shock revival in situations where the neutrino heating is not sufficient \citep{takiwaki16}. 
Such rotations may be related to magnetorotational bipolar explosions with possibly higher explosion energies than standard core-collapse supernovae \citep{moiseenko06,burrows07,takiwaki09}
and could be linked to the formation of magnetars with possible consequences on long gamma-ray bursts \citep{metzger11}. 

Besides its influence on the shock dynamics, rotation may also impact the pulsar spin at birth.
\citet{blondin07a}, using a 3D adiabatic setup, showed that SASI spiral modes have the potential to spin up a neutron star born from a non-rotating progenitor.
This surprising consequence of SASI was confirmed numerically by \citet{fernandez10}, analytically by \citet{guilet14} and
experimentally by \citet{foglizzo12,foglizzo15} who reproduced the redistribution of angular momentum by a spiral mode in a shallow-water analogue of SASI.
A SASI spiral mode could spin up a neutron star born from a non-rotating progenitor to periods in the range of $50\,\rm{ms}$ to $1\,\rm{s}$ \citep{guilet14}.
However, SASI can also develop sloshing modes which do not allow for an angular momentum redistribution.
\citet[hereafter Paper I]{kazeroni16} investigated the conditions for a spiral mode to prevail and showed that its emergence requires a minimum ratio
between the initial shock radius and the proto-neutron star radius.
Regarding rotating progenitors, \citet{blondin07a} pointed out that SASI may reduce the neutron star spin compared to a simple estimate based on angular momentum conservation
during the collapse. This could even give birth to a counter-rotating neutron star if SASI redistributes enough angular momentum.
Nevertheless, it is still unclear how efficiently a neutron star could be spun up or down by a SASI spiral mode throughout the range of stellar rotation rates.

The purpose of this paper is twofold. Using a parametric study of an idealised model, we aim at better describing the influence of rotation on the shock dynamics,
focusing on both SASI and the corotation instability. Then, we propose to study the effect of spiral modes on the natal pulsar spin in order to assess to what
extent these instabilities could spin the pulsar up or down. The paper is organized as follows. The physical and numerical models are described in Section \ref{sec:methods}.
Section \ref{sec:shock} is dedicated to the interplay between the instabilities and rotation. The relationship between the iron core rotation rate and the initial pulsar spin
is examined in Section \ref{sec:spin}. The results are synthesised in Section \ref{sec:discussion} and conclusions are drawn in Section \ref{sec:conclu}.

\section{Methods}
\label{sec:methods}

The idealised setup considered in this work is dedicated to the study of SASI in its simplest form.
It is very similar to the one described in Paper I, but includes the progenitor rotation.
In short, the model is restricted to the equatorial plane of a massive star described in cylindrical geometry. This allows non-axisymmetric modes to develop despite the 2D geometry.
The accretion flow is modelled by a perfect gas with an adiabatic index $\gamma=4/3$. The flow is decelerated by a shock wave of initial radius $\rsh_0$ with
an incident Mach number $\mathcal{M}_1=5$. The matter is then accreted subsonically onto the surface of a proto-neutron star (PNS). Its radius, defined as $r_*$,
delimits the inner boundary of the domain. Heating by neutrino absorption is ignored in order to suppress the neutrino-driven convection 
for the sake of simplicity. A cooling function is used to mimic the energy losses due
to neutrino emission, with the approximation $\mathcal{L} \varpropto \rho P^{3/2}$ \citep{blondin06} where $\rho$ and $P$ respectively stand for the density and the pressure.

\begin{figure}
\centering
	\includegraphics[width=\columnwidth]{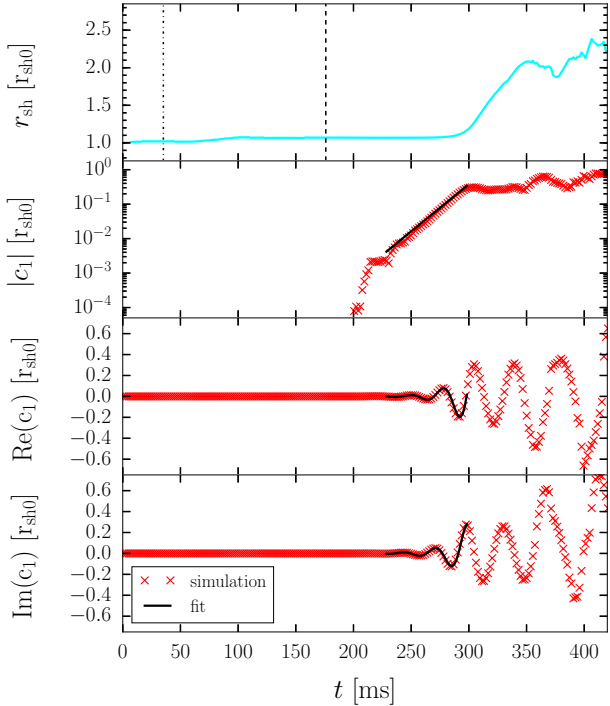}
    \caption{Time evolution of the stationary flow and early non-linear phase of the instability in the case $R=3$ and $j = 10^{16}\, \cms$. The quantities are normalised by the initial shock radius $\rsh_0$.
    \textit{Panel (a):} after the non-rotating stationary flow has relaxed on the grid for a few tens milliseconds, a constant specific angular momentum is injected through the outer boundary (dot-dashed line).
    The mean shock radius (cyan curve) increases by less than 10\%. Once the 1D rotating flow has relaxed, density perturbations are injected (dashed line).
    \textit{Panel (b):} the amplitude of the Fourier coefficient $c_1$ of the shock deformation begins to grow about an advection time after the perturbations have been injected (red crosses).
    A measurement of the oscillation frequency and the growth rate of the mode $m=1$ of SASI is performed during the well defined exponential growth of the instability in the linear regime (fit represented by the black line).
    \textit{Panels (c) and (d):} the evolution of the real part (c) and the imaginary part (d) of the Fourier coefficient (red crosses) are fitted by the function (\ref{eq:fitfunc}) during the exponential
    growth of the coefficient (black lines).
    }
    \label{fig:1Dto2D}
\end{figure}

Once the non-rotating stationary flow has relaxed on the grid for a few hundreds numerical timesteps (panel (a) of Fig. \ref{fig:1Dto2D}), 
a constant amount of specific angular momentum is injected through the outer boundary of the domain, similarly to the approach considered by \citet{iwakami14b}.
The specific angular momentum $j$, defined as $j=v_{\phi}(r)r=\Omega(r)r^2$ where $v_{\phi}$ and $\Omega$ are respectively the azimuthal velocity and the rotational velocity,
is conserved if the evolution is axisymmetric.
In this study, we consider a range of rotation rates covering three orders of magnitude: $10^{13}\, \cms \leq \textit{j} \leq 10^{16}\, \cms$.
Note that $j=10^{16}\, \cms$ would correspond to a spin period of 0.6 ms for a pulsar of 10 km radius if the angular momentum were conserved down to its surface.
According to stellar evolution calculations, the specific angular momentum contained in-between the enclosed masses $1.3\,\Msol$ and $2\,\Msol$ is such that
$j \sim 10^{14}-10^{15}\, \cms$ if the magnetic field is taken into account and $j \sim 10^{16}-10^{17}\, \cms$ otherwise \citep{heger05}.
Among the few 3D simulations including rotation, some considered very fast rotation rates such as:
$j = 1.2-6.3\times 10^{16}\, \cms$ \citep{nakamura14}, $j = 3.1\times 10^{16}\, \cms$ \citep{kuroda14} or
$j = 3.9\times 10^{16}\, \cms$ \citep{takiwaki16}. However, slower rotations have also been considered: $j = 10^{15}\, \cms$ \citep{blondin07a}, $j = 1-4\times 10^{15}\, \cms$ \citep{endeve12} and $j \leq 10^{16}\, \cms$ \citep{iwakami14b}.

We let the 1D rotating flow relax for a few more hundreds numerical timesteps (panel (a) of Fig. \ref{fig:1Dto2D})
in order to reach an equilibrium before density perturbations are introduced to trigger SASI. Rotation induces an increase of the shock radius of less than 10\% in the axisymmetric phase of the simulations.
Similarly to the approach considered in Paper I, two density perturbations at pressure equilibrium are launched through the outer boundary to initiate the two counter-rotating spiral modes $m = \pm 1$,
where $m$ is the azimuthal wavenumber of the unstable mode of SASI.
The amplitude of the perturbation is $10^{-4}$ and the asymmetry between the two counter-rotating spiral modes is $\epsilon=0.25$ (as defined by Eqs. 2-3 in Paper I).

When converting to physical units, we use the same normalisation values as in Paper I: 
a proto-neutron star with radius $r_* = 50\rm{km}$, a neutron star mass $M_*=1.3\,\Msol$ and a constant mass accretion rate $\dot{M}=0.3\,\Msol\,\rm{s^{-1}}$ 
which are typical values for the stalled shock phase of a core-collapse supernova during the first second after bounce.
As in Paper I, the parameter $R$ defines the radii ratio: $R\equiv \rsh_0/r_*$.
For non-rotating cases, the threshold $R \gtrsim 2$ is a necessary condition for a spiral mode to prevail in the non-linear regime \citep{kazeroni16}.
Below this ratio, a sloshing mode dominates the dynamics even if a spiral mode was artificially initiated in the linear regime of SASI.
In the present study, we consider the following radii ratios : $R=\{1.67,\,2,\,2.5,\,3,\,4,\,5\}$. 
In total, 140 simulations are performed to cover the parameter space.

As in Paper I, the numerical simulations are performed with a version of the RAMSES code \citep{teyssier02, fromang06} adapted to cylindrical coordinates.
RAMSES is a second-order finite volume code and employs the MUSCL-Hancock scheme. We use the HLLD Riemann solver \citep{miyoshi05} and the monotonized central slope limiter.
Except for the introduction of rotation, the numerical setup is the same as in Paper I. 
Periodic boundary conditions are used in the azimuthal direction while a reflecting condition is employed at the inner boundary, $r=r_*$, and a constant inflow set by
the 1D stationary flow at the outer boundary, $r=r_{\rm out}$.
The only modification to the previous setup concerns the numerical resolution employed.
A uniformly spaced grid is used with 1600 cells in the azimuthal direction.
The number of radial cells is 150 below the shock for $R=3$ and the grid cell size is maintained constant when $R$ varies.
The outer boundary is located at $r_{\rm out} = 6-10\ \rsh_0$. This represents a total of 1600 to 2730 cells in the radial direction.
The value of $r_{\rm out}$ is set so that the shock wave is always contained within the simulation grid even for the high rotation cases where it expands significantly (see Section \ref{subsec:shock_2}).

\section{Influence of rotation on the shock wave dynamics}
\label{sec:shock}

\subsection{Overlapping instabilities}
\label{subsec:shock_1}

The accretion flow described in the previous section is unstable to SASI when rotation is neglected (Paper I).
The centrifugal force $\Omega(r)^2r$ is not dominant in the dynamics for the range of rotation rates considered in this study. 
The advection time from the shock radius to the proto-neutron star surface is only increased by 30\% over the range of rotation rates considered in our study, while the shock radius expands by less than 10\%.
The centrifugal force is quadratic with respect to the rotation rate and remains much smaller than gravity $|-G\mns/r^2|$:
\begin{equation}
 \label{eq:fracForces}
 \frac{j^2/r^3}{G\mns/r^2} = \frac{2T}{W} = 1.2\times 10^{-1} \left(\frac{j}{10^{16}\, \cms}\right)^2\left(\frac{50\,\rm{km}}{r}\right)\left(\frac{1.3\,\Msol}{\mns}\right).
 \end{equation}
where T and W respectively stand for the rotational kinetic energy and the gravitational potential energy.
 
In order to estimate the growth rates $\omega_i$ and oscillation frequencies $\omega_r$ of SASI from our simulations, an azimuthal Fourier decomposition of the shock deformation is computed. The time evolution of the Fourier coefficients $c_{\rm m}$
is then fitted during the linear phase of SASI using a function of the form
\begin{equation}
\label{eq:fitfunc}
 f(t) = A\cos{\left(\omega_r t+\Phi\right)}\exp{\left(\omega_i t\right)},
\end{equation}
where $A$ and $\Phi$ represent the amplitude and the phase (see Fig. \ref{fig:1Dto2D} for an example of this method).

\begin{figure}
\centering
	\includegraphics[width=\columnwidth]{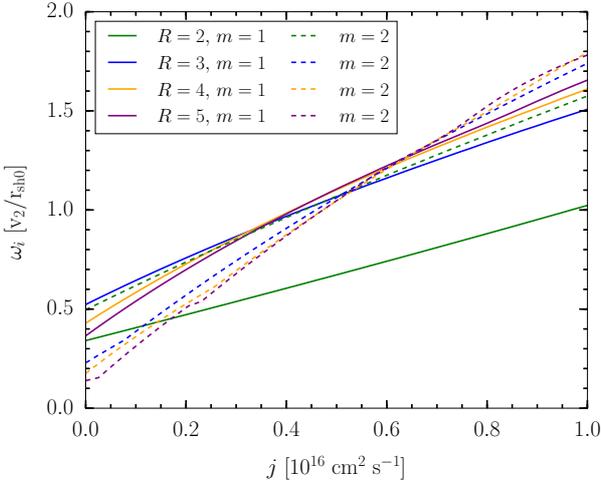}
    \caption{
    Growth rates of the most unstable modes as a function of the specific angular momentum for different values of $R$.
    The growth rate is normalised by the post-shock velocity divided by the shock radius.
    Solid lines are for the spiral modes $m=1$ and dashed lines for the spiral modes $m=2$.
    }
    \label{fig:growthrates}
\end{figure}

\begin{figure}
\centering
	\includegraphics[width=\columnwidth]{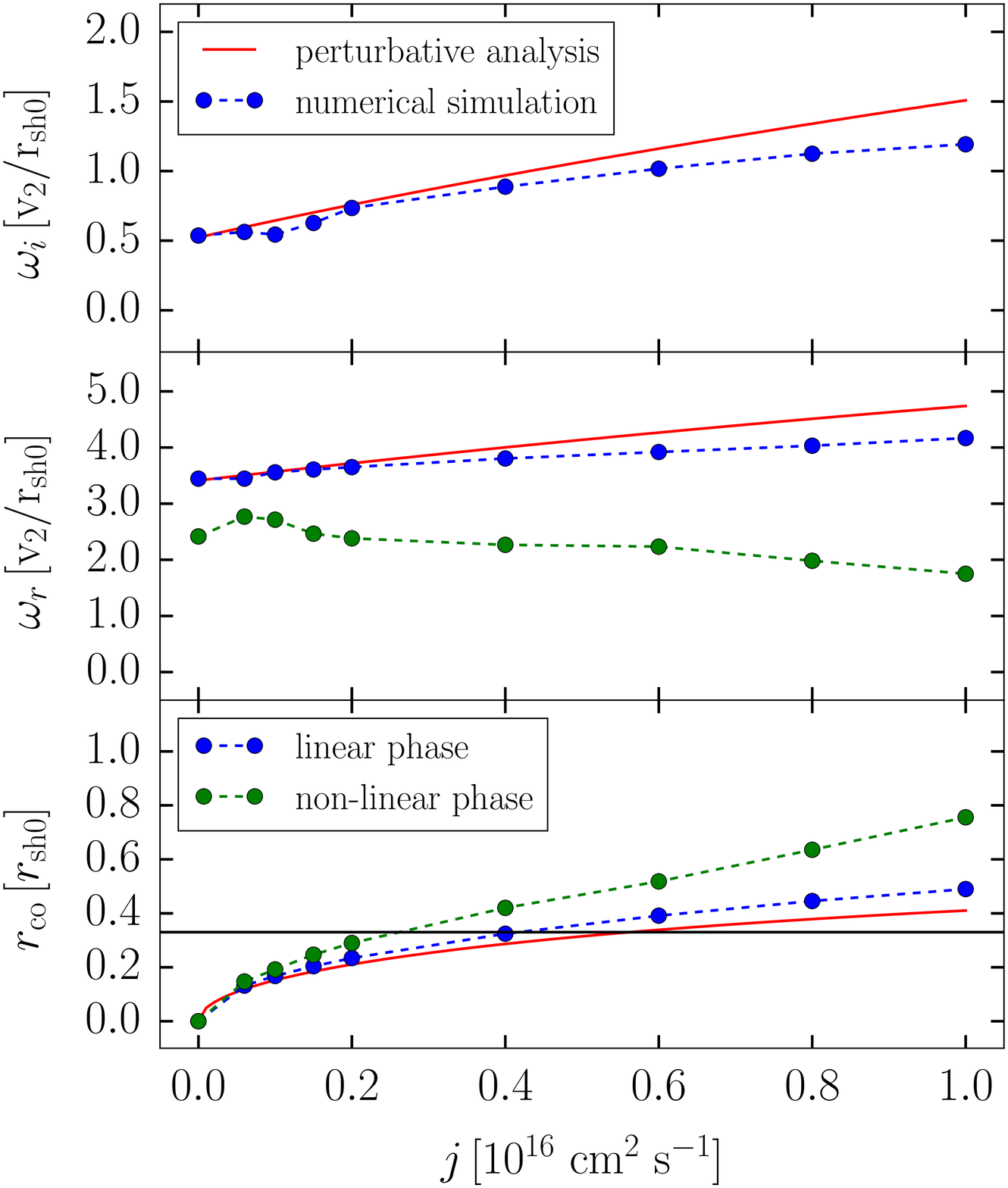}
    \caption{Impact of rotation on the linear and the non-linear regimes of the most unstable mode of SASI ($m=1$) for $R=3$.
    \textit{Upper panel:} The growth rate of the spiral mode depends linearly on $j$ as shown both by a perturbative analysis (red curve) and
    simulations of the linear regime (blue curve).    
    The values are normalised by the post-shock velocity divided by the shock radius.
    \textit{Middle panel:} The oscillation frequency increases linearly with $j$ as shown by a perturbative analysis (red curve).
    This trend is similarly obtained from numerical simulations of the linear regime (blue curve).
    In the non-linear regime, the oscillation frequency decreases with $j$ due to the expansion of the shock wave (green curve).
    \textit{Lower panel:} The corotation radius increases with $j$ as shown both by a perturbative analysis (red curve) and
    simulations (blue curve). It is slightly higher in the non-linear phase (green curve) than in the linear regime.
    The black line represents the PNS radius below which no corotation exists. The values are normalised by the initial shock radius.}
    \label{fig:lin}
\end{figure}

\begin{figure*}
 \centering
\includegraphics[width=0.9\columnwidth]{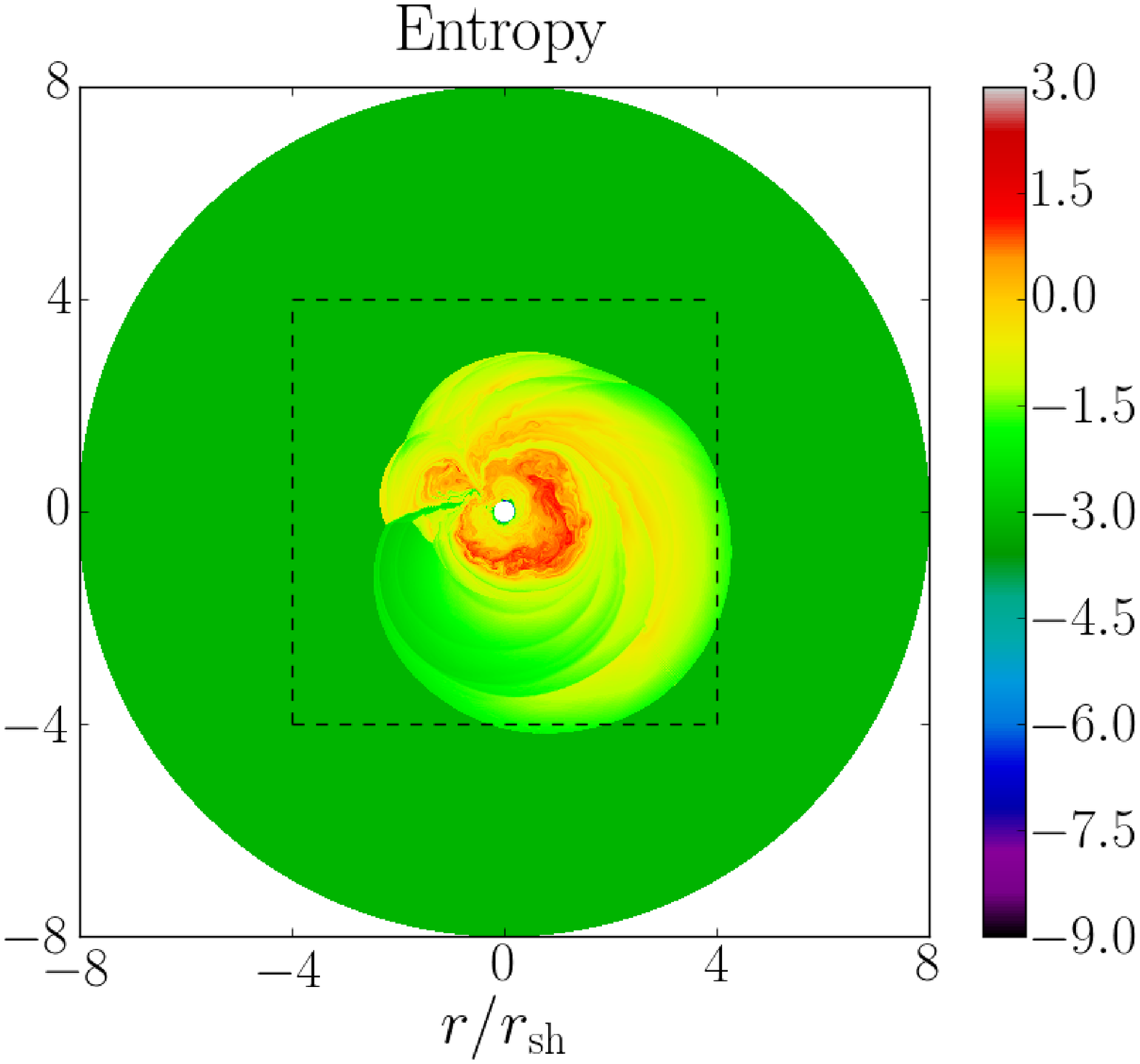}
\includegraphics[width=0.9\columnwidth]{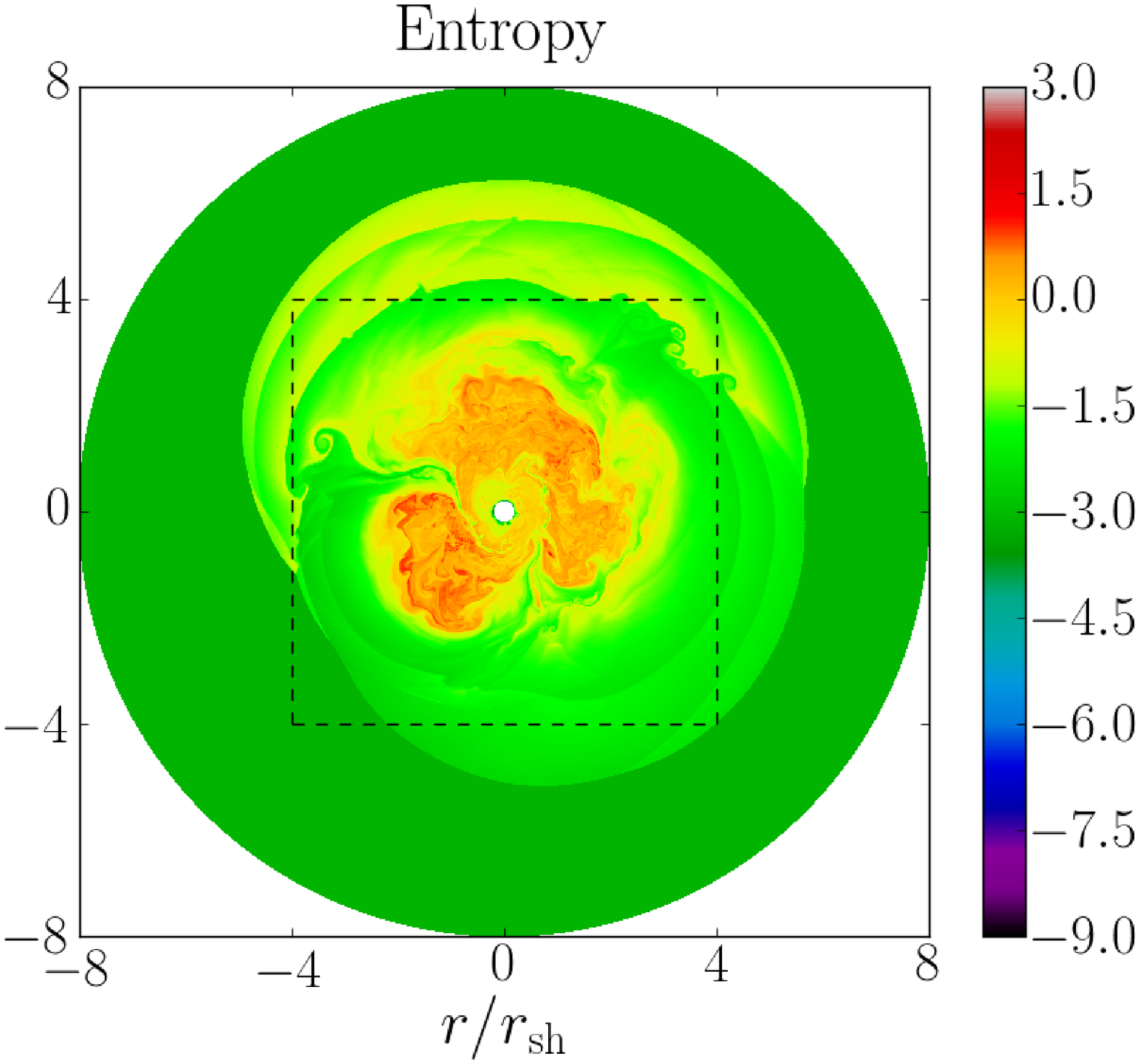}
\includegraphics[width=0.9\columnwidth]{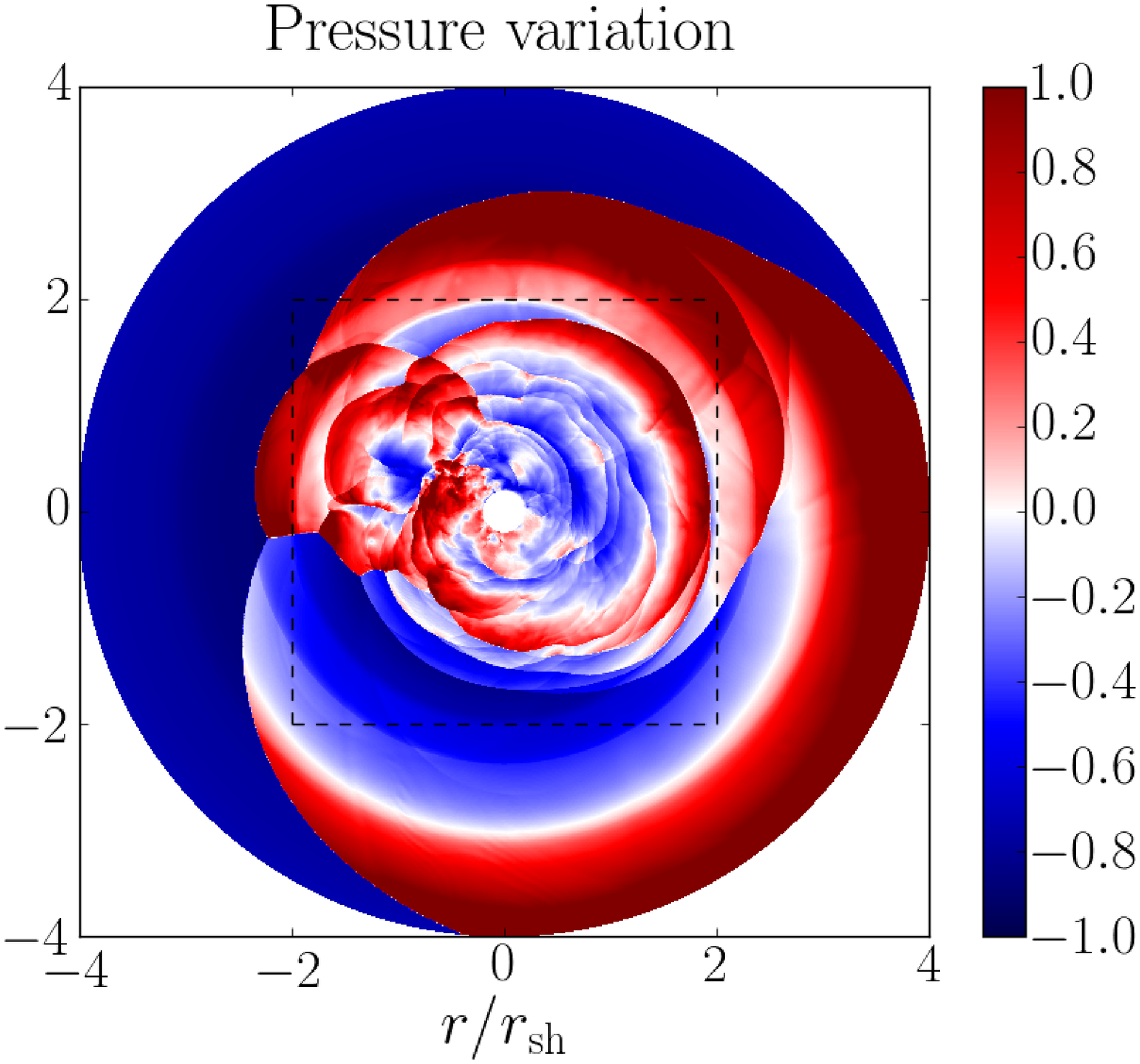}
\includegraphics[width=0.9\columnwidth]{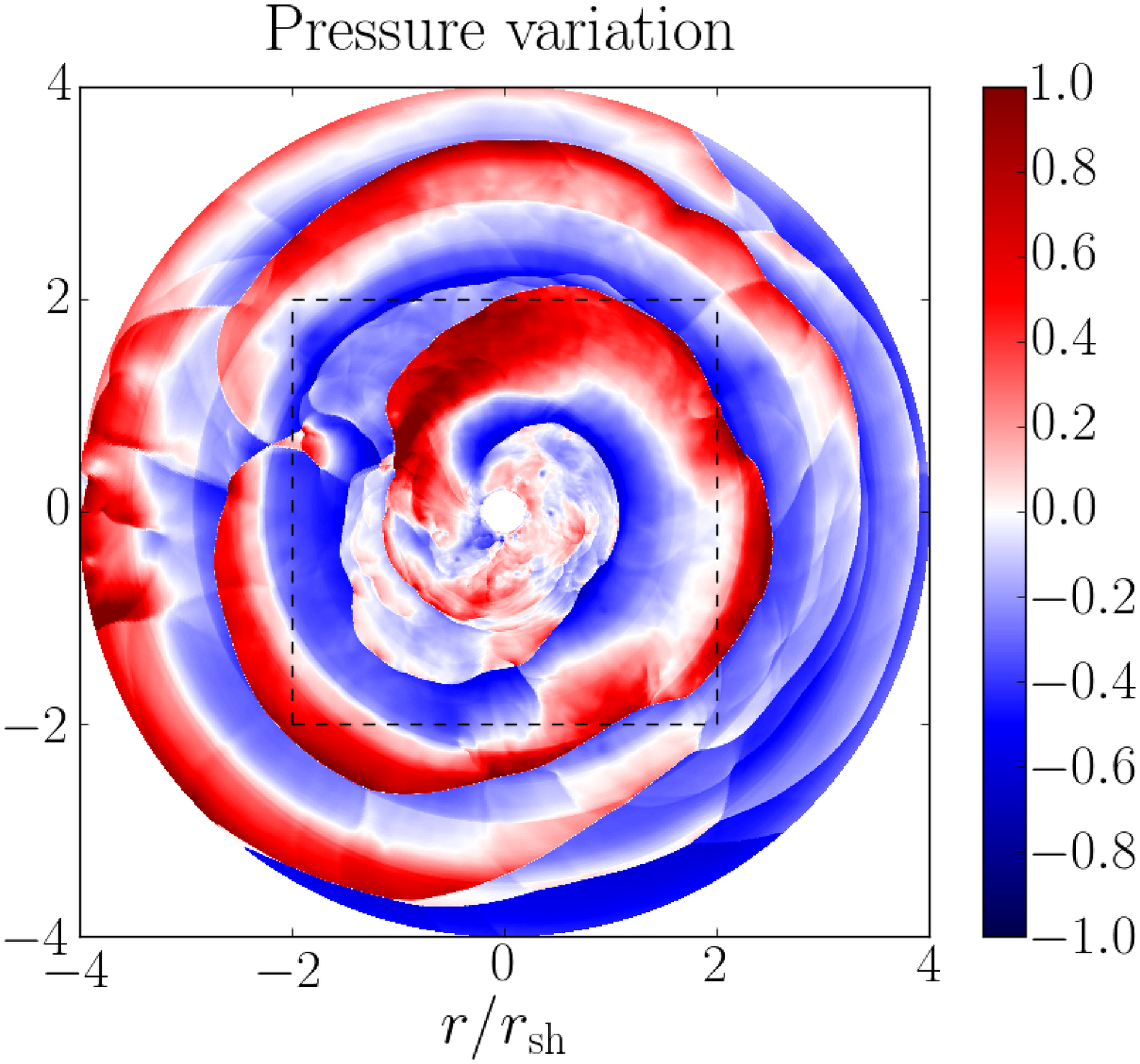}
\includegraphics[width=0.9\columnwidth]{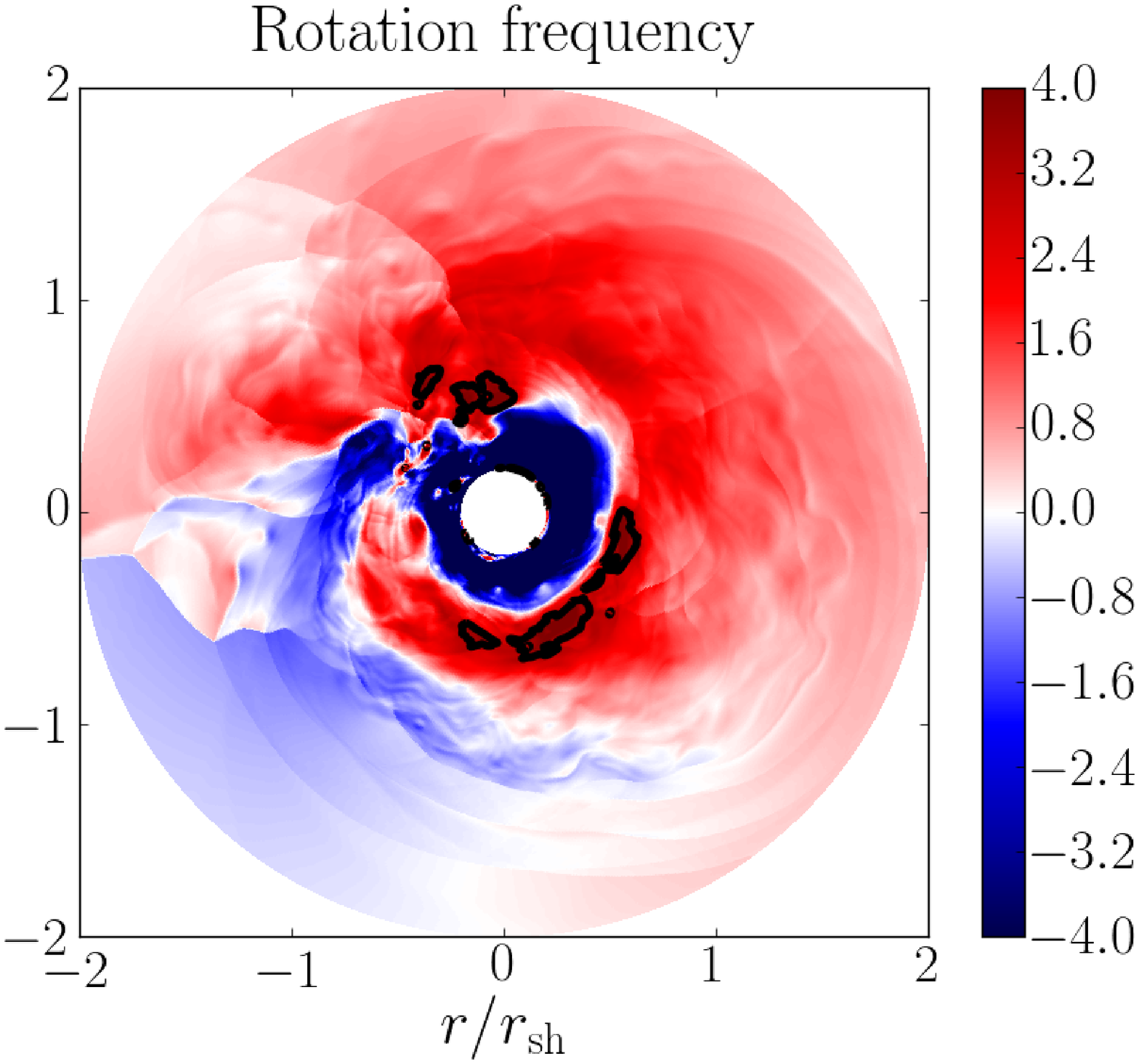}
\includegraphics[width=0.9\columnwidth]{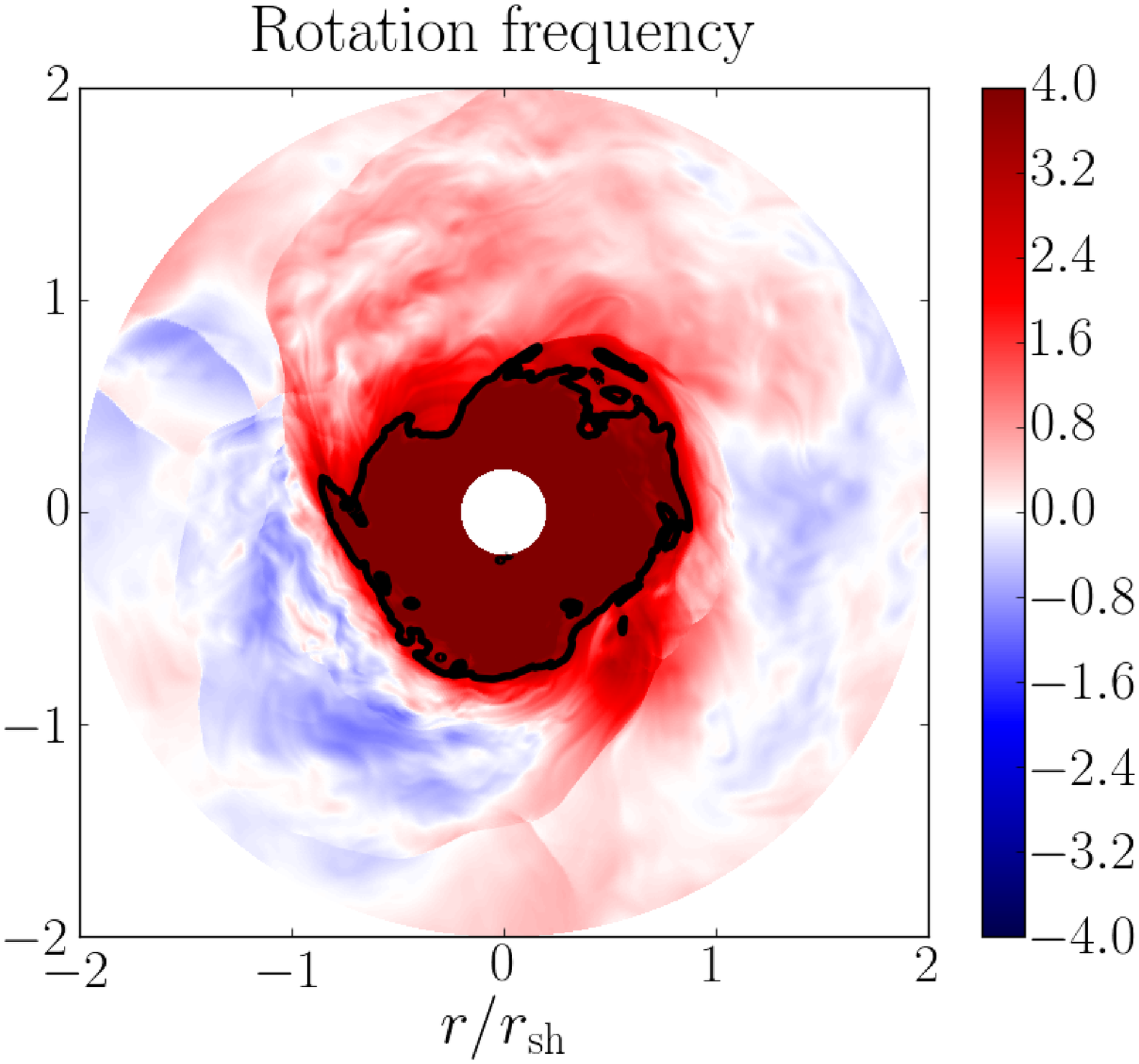}
  \caption{
  A SASI-dominated case with $R=5$ and $j=10^{15}\, \cms$ (left panels) is compared to a case dominated by a corotation instability with $R=5$ and $j=6\times 10^{15}\, \cms$ (right panels).
  \textit{Upper panels:} entropy snapshots of the non-linear dynamics. Both cases are dominated by a spiral mode $m=1$ and
  a greater shock expansion is reached in the faster rotating case. The black dashed lines delimit the region zoomed in the snapshots below.
  \textit{Middle panels:} pressure variation with respect to the azimuthal averaged value computed as $\left(P-\langle P\rangle_{\phi}\right)/P$.
  A tightly-wound spiral wave is observed only is the rotation if fast enough. The black dashed lines delimit the region zoomed in the snapshots below.
  \textit{Lower panels:} rotation frequency defined as $v_{\phi}/r$. The black lines indicate a corotation radius that delimits the regions in which the flow has a higher rotation frequency than the spiral mode.
  A well defined corotation develops only for high enough rotation rates and could trigger a second instability overlapping with SASI. The corotation instability seems to generate a strong open one-armed spiral wave.}
\label{fig:spiral}
  \end{figure*}

The perturbative study of \citet{yamasaki08} revealed that the linear effect of the rotation on the growth rate of SASI is due to a Doppler shift of the frequencies. 
This shift induces a linear increase of the growth rate of prograde spiral modes with respect to the specific angular momentum (Fig. \ref{fig:growthrates}).
Our set of simulations confirms the linear increase of the growth rate as shown in the case R = 3 (Fig. \ref{fig:lin}, upper panel), and also demonstrates a linear increase of
the oscillation frequency (Fig. \ref{fig:lin}, middle panel). 
Recently, \citet{blondin17} obtained similar results in their simulations and showed that the linear trends hold for the 2D cylindrical geometry as well as the 2D and 3D spherical geometries.
The comparison with the perturbative analysis shows a 8\% discrepancy of growth rates at low rotation rate
and up to 15\% at high rotation rate. The discrepancy for oscillation frequencies is significantly smaller, less than 1\% at low rotation and less than 10\% at high
rotation. The increase of the discrepancy with rotation is not explained. It is reminiscent of the discrepancy observed by \citet{foglizzo07} in the simulations
of \citet{blondin06}, which was also largest for the largest shock distance. Note that \citet{blondin17} also observed a departure from the linear trend at high rotation rates.

A corotation radius $r_{\rm{co}}$ is defined by:
\begin{equation}
 \label{eq:corot}
 r_{\rm{co}} \equiv \sqrt{\frac{mj}{\omega_r}}.
\end{equation}
This delimits a region inside which the flow rotates faster than the spiral pattern.
Such a region can only exist if the rotation rate is high enough (Fig. \ref{fig:lin}, lower panel).
Note that a corotation radius can emerge in the non-linear regime although it does not exist in the linear regime. 
This is a consequence of a wider post-shock region in the non-linear regime which increases the advection time and diminishes the oscillation frequency (Fig. \ref{fig:lin}, middle panel).
The oscillation frequency in the non-linear regime is estimated by computing the rotation frequency of the triple point.
For $R=3$, a corotation radius can appear with a specific angular momentum 30\% lower than predicted by the linear analysis.
Its emergence above the inner boundary, located at $r=r_*$, does not seem to affect the linear increase of the growth rates and the oscillation frequencies (Fig. \ref{fig:lin}).

\citet{takiwaki16} identified an explosion aided by a corotation instability in a simulation of a fast rotating case ($j = 3.9 \times 10^{16}\, \cms$) where
a strong open one-armed spiral wave generates turbulent kinetic energy which complements the energy deposition by neutrinos and successfully revives the stalled shock.
Such a situation seems to occur only if the dynamics is SASI-dominated as in the $27\,\Msol$ progenitor of their study.
The mechanism responsible for SASI has been well explained as an advective-acoustic cycle \citep{foglizzo07,guilet12}. By contrast, the origin of the corotation instability remains uncertain.
In the different context of an isolated neutron star rotating differentially,
it has been proposed that this instability is generated by an energy transfer occurring across a band around a corotation radius where acoustic waves could be amplified \citep{watts05,passamonti15}.
Numerically, it has been observed that the corotation instability can develop when the ratio T/|W| exceeds a value of the order of 1\%
both in differentially rotating neutron stars \citep{shibata02,shibata03} and in rotating collapsing stellar cores \citep{ott05,kuroda14}. 
Despite quite different mechanisms from a theoretical point of view, \citet{kuroda14} could not clearly distinguish SASI from the corotation instability in their simulations of a $15\,\Msol$ progenitor. 

Although a spiral mode $m=1$ always dominates the non-linear shock dynamics in our simulations (Fig. \ref{fig:spiral}, upper panels),
a tightly-wound spiral flow is observed in the post-shock region at large enough rotation rate (Fig. \ref{fig:spiral}, middle right panel) and resembles the open one-armed spiral wave identified by \citet{takiwaki16}.
Such a flow pattern is never witnessed in cases where a corotation radius cannot be well defined (Fig. \ref{fig:spiral}, lower panels), that is to say where SASI seems to be the only possible instability.
This open one-armed spiral pattern is thus suggestive of a distinct instability. It is however not clear yet whether such a feature is a systematic signature of the corotation instability.

\subsection{Shock evolution}
\label{subsec:shock_2}

\begin{figure}
\centering
	\includegraphics[width=\columnwidth]{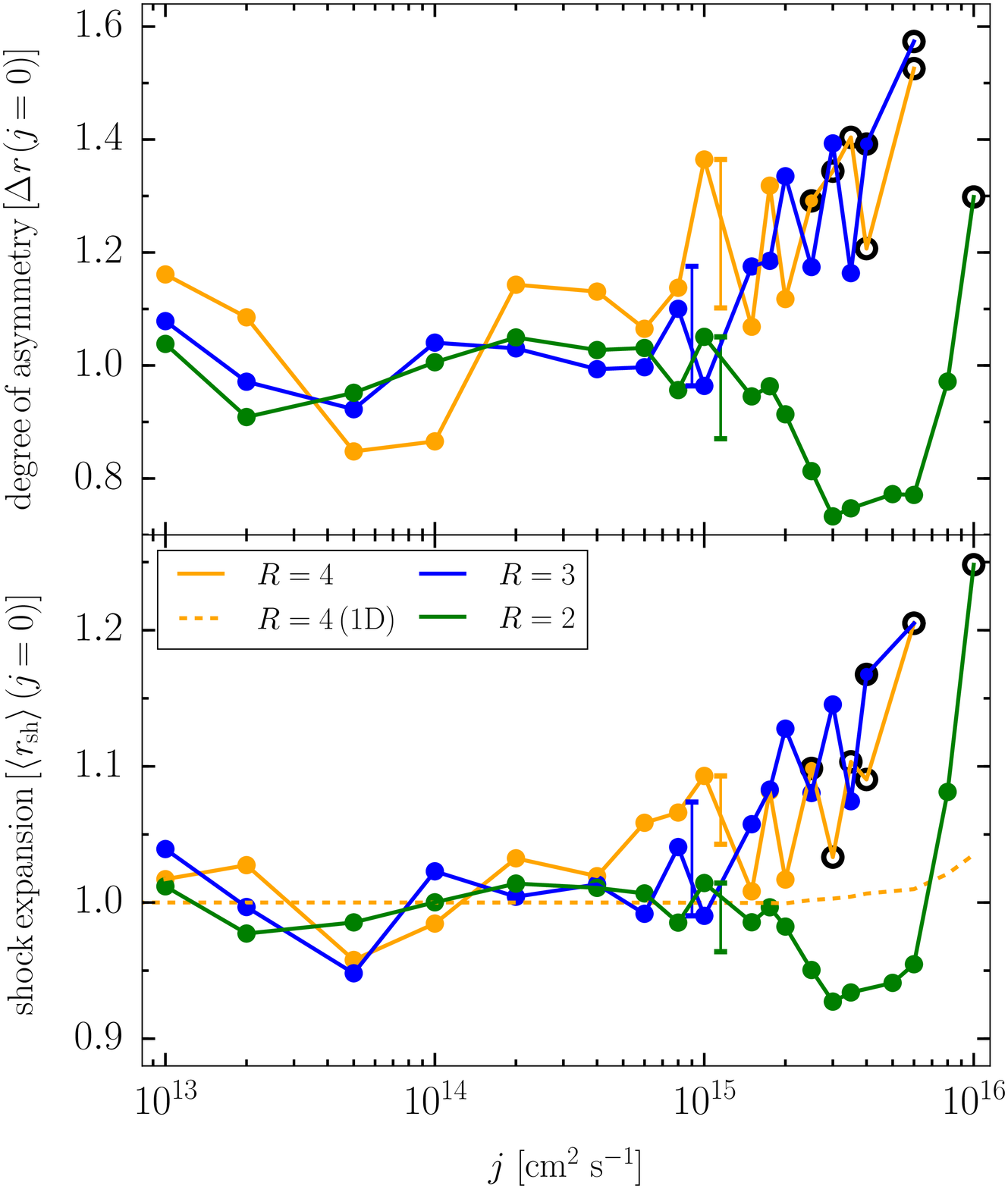}
    \caption{Influence of rotation on the non-linear regime of SASI for different values of $R$. 
    \textit{Upper panel:} saturation amplitude of the mode $m=1$ as a function of $j$.
    \textit{Lower panel:} mean shock radius as a function of $j$.
    The dashed curve shows the effect of the centrifugal force in 1D simulations.
    The black circled points denote that a corotation emerges in the linear regime (empty circles) or in the non-linear regime (full circles).
    In both panels, the values are averaged in the non-linear regime and normalised by the ones measured in the non-rotating cases.
    To reduce the variability of the results, the non-rotating simulations are repeated five times with slightly different initial perturbation amplitudes.
    The simulations where $j = 10^{15}\, \cms$ are also repeated five times to assess the stochasticity of our results. The vertical bars indicate the minimum and maximum values obtained for each value of $R$.
    }
    \label{fig:shock_SASI}
\end{figure}

\begin{figure}
\centering
	\includegraphics[width=\columnwidth]{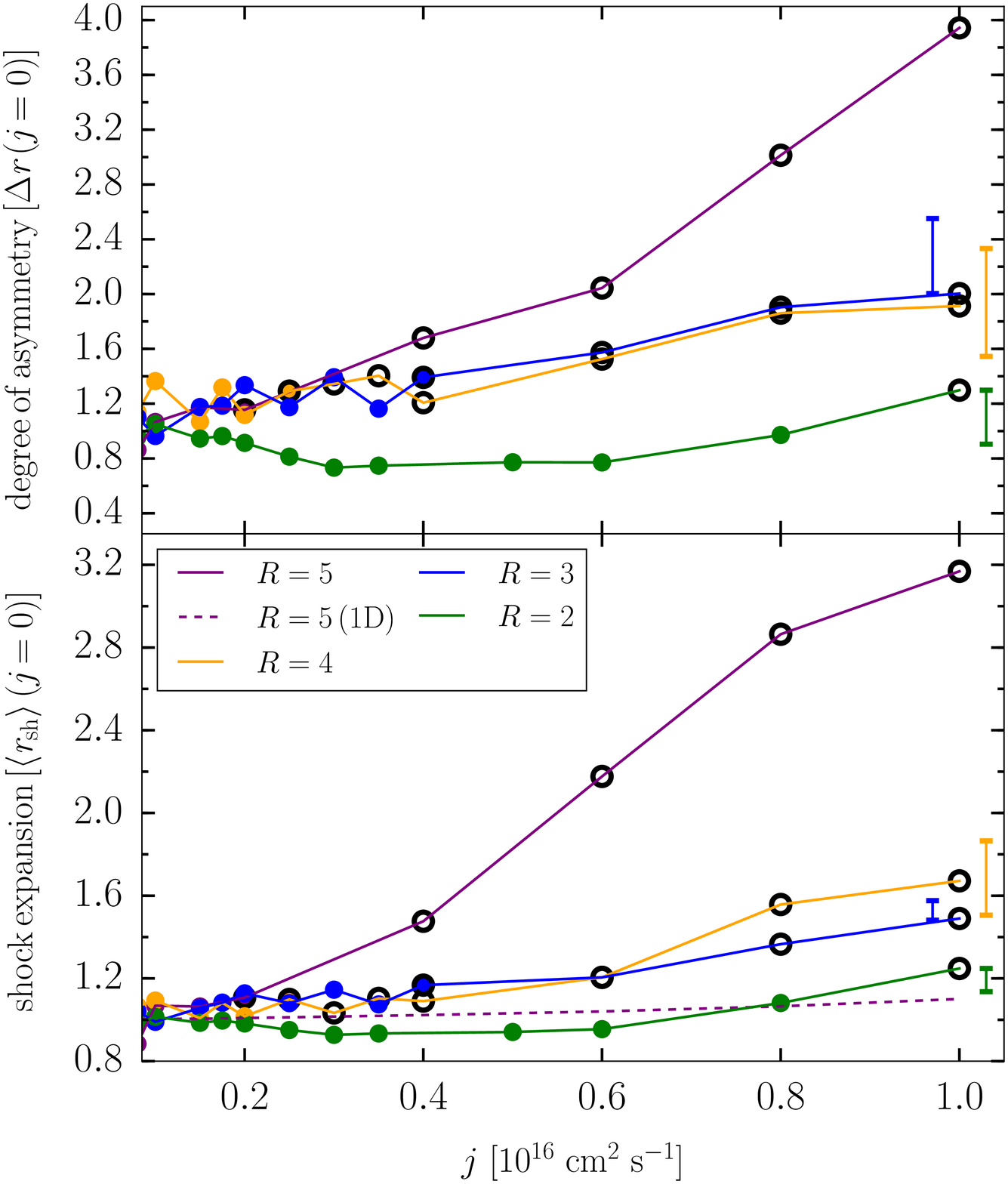}
    \caption{
    Same as in Fig. \ref{fig:shock_SASI} for the corotation instability.
    In this case, the simulations where $j = 10^{16}\, \cms$ are repeated five times.
    }
    \label{fig:shock_TW}
\end{figure}

In this section, we focus on the influence of rotation on the shock wave dynamics.
For this purpose, the mean radius $\langle\rsh\rangle$ of the shock and its degree of asymmetry $\Delta r$, defined as the dominant coefficient of the Fourier transform of the shock deformation,
are time-averaged over the non-linear regime of the simulations.
The degree of asymmetry corresponds to the saturation amplitude of the dominant mode of SASI and may set the magnitude of the neutron star kick \citep{scheck04,scheck06}. 
Fig. \ref{fig:shock_SASI} shows the influence of rotation on the dynamics compared to non-rotating cases in situations where SASI is the only instability involved.
For each value of $R$, five simulations of the non-rotating case are performed in order to estimate the variability of the results, as discussed later in this section.
These simulations differ only by a minor change of the initial perturbation amplitude.
Both the shock expansion and the degree of asymmetry are non monotonic functions of the rotation rate. In fact, these quantities increase with the angular momentum only
if the condition $R>2$ is fulfilled. In the absence of a corotation radius and if $R>2$, rotation increases the shock expansion and
the degree of asymmetry of the SASI spiral mode by respectively 15\% and 30\%. 
The larger the radii ratio, the lower the angular momentum threshold for rotational effects on SASI.

Below the ratio $R=2$, it was observed in Paper I that sloshing modes dominate the non-linear regime of SASI when rotation is neglected.
For $R=2$, a minimal specific angular momentum of $j=10^{15}\, \cms$
is necessary to destabilise a sloshing mode into a spiral mode. Surprisingly, the shock expansion and the degree of asymmetry respectively
diminish by 10\% and 20\% between $j=10^{15}\, \cms$ and $j=3\times 10^{15}\, \cms$.  
This decrease is related to the non-linear saturation amplitude since the growth rates of the most unstable modes are still increasing functions of the rotation rate (Fig. \ref{fig:growthrates}).

Interestingly, the critical radii ratio required for the rotation to increase the degree of asymmetry is similar to the ratio $R=2$ identified in Paper I.
We suggest that this may not be a coincidence and we rather propose a mechanism for the symmetry breaking between counter-rotating spiral modes, which can explain the equality of these critical ratios.
For this purpose we will assume that the rotation induced by a spiral mode has a similar effect on its saturation amplitude as the rotation of the progenitor. 
For $R>2$, if a spiral mode has a slightly larger amplitude than the counter-rotating spiral, the rotation it induces will cause it to grow to even larger amplitudes until it dominates the dynamics. 
On the other hand for $R<2$, the induced rotation would cause the amplitude of the spiral to diminish and therefore tend towards a state where the two counter-rotating spirals have the same amplitude, i.e. a sloshing mode.

These results suggest that rotation might not always ease the shock revival in ab-initio 3D simulations of core-collapse supernovae. On the one hand, if $R >2$ rotation enhances the spiral mode of SASI which may facilitate the explosion \citep{fernandez15}.
On the other hand, for low radii ratios rotation reduces the amplitude of the spiral mode, disfavouring the shock revival.
To assess the stochasticity of our results, we performed five different realisations of each simulation where $j = 10^{15}\, \cms$, varying the initial perturbation amplitude by a few percent.
The vertical bars in the Fig. \ref{fig:shock_SASI} indicate that the shock expansion and the degree of asymmetry can vary respectively by 10\% and 20\%.
These fluctuations are as large as the effect of rotation on the shock dynamics. Thus a single ab-initio 3D simulation including rotation may not be sufficient to assess the impact of rotation on SASI and the explosion
due to the stochastic nature of the non-linear regime of SASI.
The consequences of the stochasticity on the explodability were pointed out by \citet{cardall15} who showed that a critical neutrino luminosity cannot be well defined for SASI-dominated cases. 
The same conclusion may be valid for rotation, whose impact on the dynamics may be altered by stochasticity.

Finally, the presence of a corotation radius greatly impacts the dynamics as shown in Fig. \ref{fig:shock_TW}.
The shock expansion and the degree of asymmetry increase more steeply when SASI overlaps with a corotation instability.
This second instability increases the mean shock radius and the saturation amplitude by a factor up to respectively 3 and 4 for $R=5$, compared to non-rotating progenitors.
For cases where $R\leq 2$, the corotation instability modifies the influence of rotation on the dynamics. While the degree of asymmetry decreases with rotation when
the corotation is absent, it increases when the corotation radius exceeds the PNS radius.
The impact of corotation is more important with greater values of $R$ because the frequency of the fundamental mode is decreased and
the minimal rotation rate for a corotation to emerge is thus reduced.
The amplification of the spiral mode by the corotation instability is larger than the stochastic fluctuations as illustrated by the vertical bars in Fig. \ref{fig:shock_TW}.
The spiral mode becomes more vigorous in the presence of a corotation instability and this may have consequences on the explosion mechanism \citep{takiwaki16}.
The diversity of the effects of rotation on the shock dynamics is summarised in table \ref{tab:tab1}. 
\begin{table}
\begin{center}
   \begin{tabular}{| l | c | c | }
  \hline
   & $\langle \rsh\rangle/r_{\rm sh0}$ & $\Delta r /r_{\rm sh0}$ \\ \hline
  centrifugal force (1D) & $1.05$ & $-$ \\ \hline
  SASI without rotation & $1.3-2.3$ & $30-60\%$ \\ \hline
  SASI without a corotation & $1.4-2.8$ & $40-80\%$ \\ \hline
  SASI with a corotation & $2-4$ & $60-200\%$ \\ \hline
\end{tabular}
\end{center}
 \caption{This table summarises the properties of the dynamics depending on the regime considered.
 The shock radius (second column) and the degree of asymmetry (third column) are normalised by the initial shock radius $r_{\rm sh0}$.
 From top to bottom, the different regimes are: centrifugal force studied in 1D simulations,
 SASI in non-rotating progenitors (see Paper I), SASI without a corotation instability
 and SASI with an overlapping corotation instability.}
\label{tab:tab1}
\end{table}

\section{Pulsar spin}
\label{sec:spin}

\subsection{Pulsar spin estimate}
\label{subsec:spin_1}

In Paper I, we showed that the spin periods measured in our simulations of a cylindrical setup agree relatively well with the analytical estimate of \citet{guilet14} adapted to this geometry.
These results are based on the idealised scenario proposed by \citet{blondin07a} in which the mass cut radius coincides with the surface where the angular momentum changes sign.
In that study, the derivation of the pulsar spin neglected the angular momentum of the iron core collapsed into a proto-neutron star and the parameter space was restricted to a single rotation rate.
In order to clarify the issue of a pulsar spin-up or spin-down, we propose to disentangle the respective contributions of the angular momentum contained initially in the core and the quantity redistributed by a spiral mode.

Considering a spiral mode in a rotating iron core, the total angular momentum of a neutron star at birth, noted $L_{\rm NS}$, can be estimated by:
\begin{equation}
 \label{eq:Lns}
  L_{\rm{NS}} \equiv L_{\rm{CORE}} - \Delta L.
\end{equation}
where $\Delta L$ corresponds to the angular momentum redistributed by a spiral wave and $L_{\rm CORE}$ to the quantity contained in the core prior to collapse.
The angular momentum redistribution can be evaluated by:
\begin{equation}
\label{eq:DL}
  \Delta L \equiv L_{\rm{SPIRAL}} - L_{\rm{0}},
\end{equation}
where $L_0$ represents the angular momentum contained in the stationary flow and
$L_{\rm{SPIRAL}}$ the quantity stored below the shock wave.
As in Paper I, we will assume that the quantity opposite to $L_{\rm SPIRAL}$ is accreted onto the PNS. 
To evaluate this quantity, we first compute the mean radial profile of the angular momentum density:
\begin{equation}
 \label{eq:lz}
 l_z\left(r,t\right) \equiv Hr^2 \int_{0}^{2\pi} \rho v_{\phi}d\phi,
\end{equation}
where $H$ is an arbitrary parameter which corresponds to the height of the cylinder. The results are independent of its value, which can be set so that $\dot{M}=0.3\,\Msol\,\rm{s^{-1}}$ .
$L_{\rm SPIRAL}$ is computed using the following formula:
 \begin{equation}
  \label{eq:Lspi}
  L_{\rm{SPIRAL}} \equiv \frac{1}{T}\int_{t_0}^{t_0+T}\int_{r_{\rm{cut}}}^{\langle \rsh \rangle} l_z(r,t)\,dr\,dt,
 \end{equation}
where $t_0$ corresponds to the initial time of the non-linear regime, $T$ to the time interval considered and $r_{\rm{cut}}$ to the mass cut radius. This radius is defined
by computing the radius above which the mean radial profile of the angular momentum density in the non-linear regime is greater than the initial one.
This method gives an upper bound on the amount of angular momentum redistributed by a spiral mode.

The inner boundary of our model does not include the neutron star. 
An additional assumption on the PNS interior is needed to evaluate the quantity $L_{\rm CORE}$.
The PNS is assumed to result from the contraction of the inner core ($1.3\,\Msol$),
which roughly corresponds to a mass coordinate of 1500 km (see \eg \citealt{marek09}), into a 50 km radius object.
We will consider idealised density and angular velocity profiles of the inner 1500 km region. The density profile is defined by:
  \begin{equation}
  \label{eq:rho_core}
    \rho_{\rm{CORE}}(r)=
    \begin{cases}
      \rho_0 & \text{if}\ r<r_{\rm{int}}, \\
      \rho_0\left(\frac{r_{\rm{int}}}{r}\right)^3 & \text{if}\ r\geq r_{\rm{int}},
    \end{cases}
  \end{equation}
with $r_{\rm{int}}=300\,\rm{km}$. For $\rho_0=4\times10^9\,\rm{g.cm^{-3}}$, the mass contained in the sphere of 1500 km radius is about $1.3\,\Msol$.
This profile is chosen to mimic more realistic conditions while being consistent with the use of a constant mass accretion rate at the outer boundary.
The angular velocity profile is defined by:
  \begin{equation}
  \label{eq:omega_core}
    \Omega_{\rm{CORE}}(r)=
    \begin{cases}
      \Omega_0 & \text{if}\ r<r_0, \\
      \Omega_0\left(\frac{r_0}{r}\right)^2 & \text{if}\ r\geq r_{0},
    \end{cases}
  \end{equation}
where $r_0=1000\,\rm{km}$ and $\Omega_0=j/r_0^2$. Such a profile is intended to imitate the results of stellar evolution calculations 
(see \eg \citealt{ott06}) while being consistent with the assumption of a constant specific angular momentum injected at the outer boundary.
The angular momentum of the core, which would be conserved during an axisymmetric collapse, can be computed as follows:
\begin{equation}
 \label{eq:Lcore}
 L_{\rm{CORE}} \equiv 4\pi \int_0^{1500\,\rm{km}} \rho_{\rm CORE}(r)r^2\Omega_{\rm CORE}(r)r^2\,dr.
\end{equation}

\subsection{Angular momentum redistribution in the post-shock region}
\label{subsec:spin_2}

\begin{figure}
\centering
	\includegraphics[width=0.8\columnwidth]{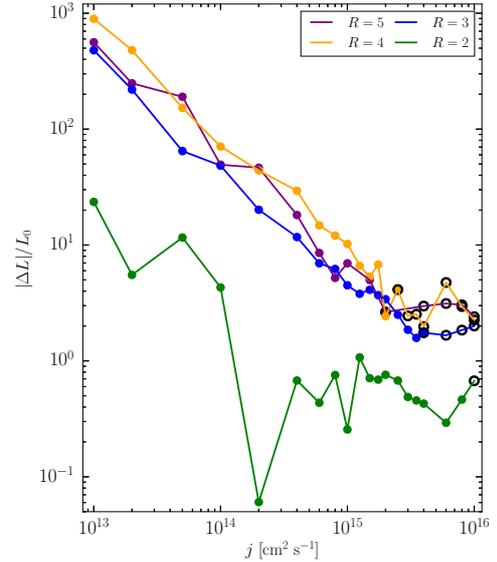}
    \caption{Angular momentum redistribution in the post-shock flow, defined by the ratio $\Delta L/L_0$ for different values of $R$.
    The black circled points denote that a corotation emerges in the linear regime (empty circles) or in the non-linear regime (full circles).
    }
    \label{fig:DL_disc}
\end{figure}

The angular momentum redistribution due to a spiral mode in the post-shock flow is displayed in Fig. \ref{fig:DL_disc}.
This redistribution significantly exceeds the initial angular momentum only for slow rotation rates, i.e. $j \lesssim 10^{14}\, \cms$, and if a SASI spiral mode emerges, i.e. if $R>2$.
In this regime, the quantity $L_{\rm{SPIRAL}}$ does not depend on the angular momentum injected and equals the non-rotating case.

As the rotation rate increases, the ratio $\Delta L/L_0$ converges towards a constant value. The higher the radii ratio, the higher this 
asymptotic value. The break of slopes is induced by the corotation which sets a minimal limit to the ratio between the angular momentum redistributed and the initial quantity.
As in Paper I, we observe that high radii ratios are more efficient at redistributing angular momentum. This is a consequence of the higher saturation amplitudes
associated with high radii ratios.

\subsection{Influence of spiral modes on the pulsar spin}
\label{subsec:spin_3}

\begin{figure}
\centering
	\includegraphics[width=0.8\columnwidth]{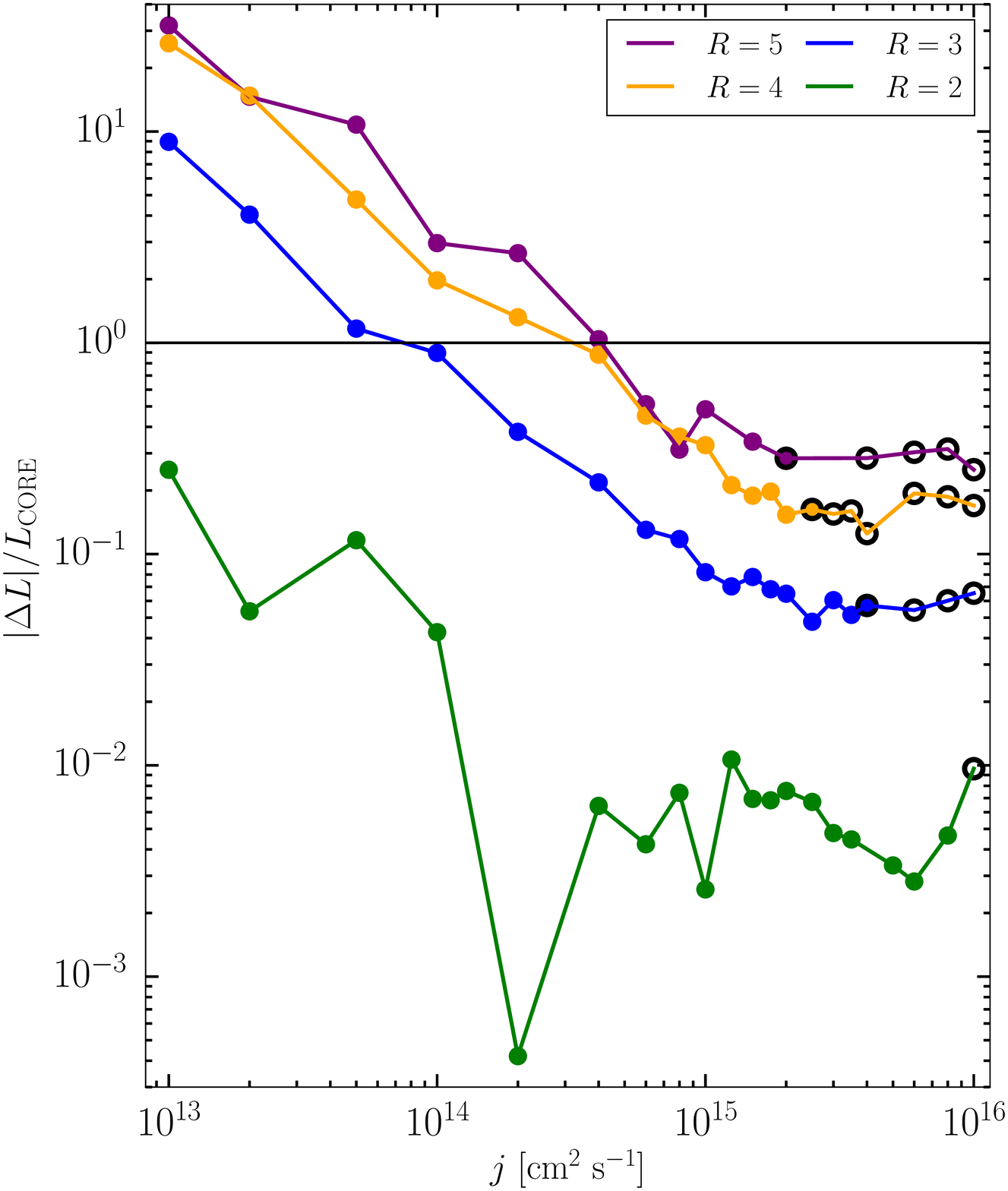}
    \caption{Ratio between the angular momentum redistributed by a spiral mode and the quantity contained in the core prior to collapse.
    Above the horizontal black line, the spiral mode mostly sets the spin period and these situations
    give birth to a counter-rotating neutron star.
    The black circled points denote that a corotation emerges in the linear regime (empty circles) or in the non-linear regime (full circles).}
    \label{fig:Ls_Lc}
\end{figure}

\begin{figure}
\centering
	\includegraphics[width=0.9\columnwidth]{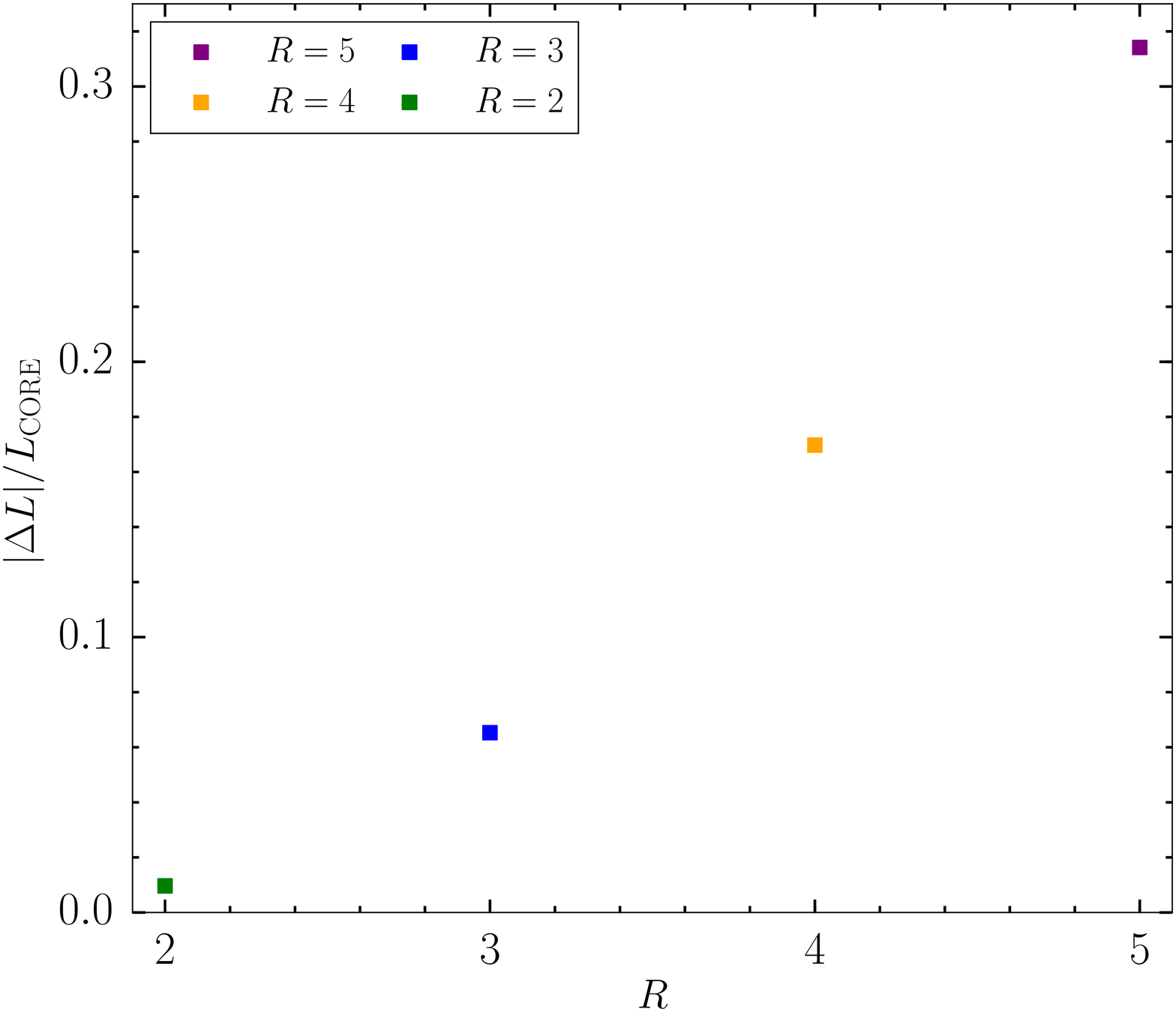}
    \caption{Average ratio between the angular momentum redistributed by the spiral mode and the quantity contained in the iron core,
    computed over cases where a corotation radius emerges.
    This ratio reflects the efficiency of the corotation instability to spin down the pulsar for different values of $R$.}
    \label{fig:Lz_TW}
\end{figure}

The importance of the spiral modes in the angular momentum budget of the collapse of a massive star depends both on the rotation rate and the radii ratio (Fig. \ref{fig:Ls_Lc}).
In cases where the angular momentum redistributed, $\Delta L$, exceeds the one contained in the core, $L_{\rm CORE}$, SASI can give birth to counter-rotating neutron stars.
This is observed for high radii ratios, $R \geq3$, and low rotation rates: $j \leq 1-5\times 10^{14}\, \cms$. In this region of the parameter space, non-axisymmetric modes are the dominant contribution to the angular momentum budget.

On the contrary, we identify two distinct regions of the parameter space where non-axisymmetric modes do not play a dominant role on the neutron star spin.
The first one corresponds to low radii ratios: $R \leq 2$. In such cases, rotation hardly affects SASI (Fig. \ref{fig:shock_SASI}) and the saturation amplitude of the spiral is too low
to drive significant angular momentum redistribution \citep{guilet14}.
The second one is related to the corotation instability. The amount of angular momentum redistributed can only account for less than 30\% of the core one (Fig. \ref{fig:Lz_TW}). 
This indicates that the corotation instability may only modestly spin down the neutron star and does not represent an efficient braking mechanism. 

Note that in the case considered by \citet{blondin07a}: $R = 5$ and $j = 10^{15}\, \cms$, the angular momentum in the core overcomes the quantity redistributed by SASI (Fig. \ref{fig:Ls_Lc})
and this does not lead to a counter-rotating neutron star if the core angular momentum is taken into account in the derivation of the neutron star spin period. Counter-rotating neutron stars
may only concern progenitors with a rotation rate at least 2 to 5 times slower.

\begin{figure}
\centering
	\includegraphics[width=\columnwidth]{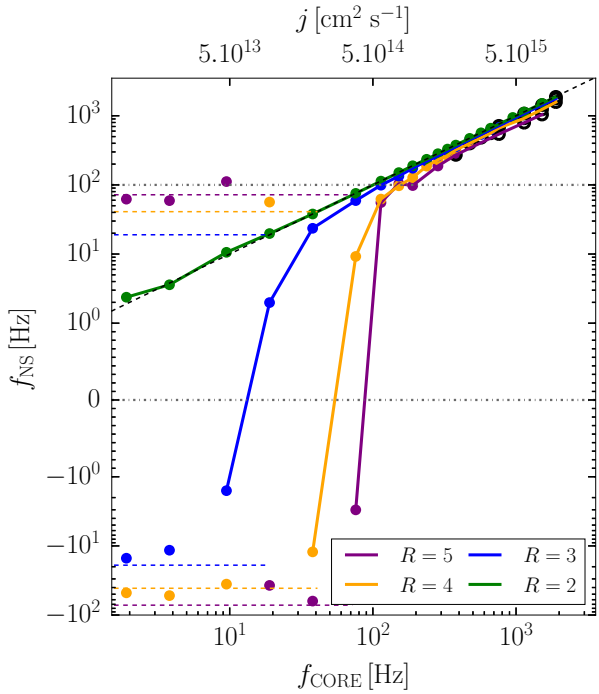}
    \caption{Relationship between the rotation frequency inferred from a spherically-symmetric collapse ($f_{\rm CORE}$) and the one which accounts for the development of spiral modes ($f_{\rm NS}$).
    The black dashed line illustrates the absence of angular momentum redistribution ($f_{\rm NS} = f_{\rm CORE}$).
    Neutron stars above and below this line are respectively spun up and down by SASI. 
    The upper grey dot-dashed line indicates the maximal rotation frequency which seems compatible with the observations ($f \lesssim 100\,\rm{Hz}$).
    Counter-rotating neutron stars correspond to the cases below the lower grey dot-dashed line.
    The horizontal dashed lines refer to the rotational frequencies obtained in simulations without rotation.
    The angular momentum redistribution at low rotation rates (isolated points) is very similar to the non-rotating cases.
    }
    \label{fig:Pns_Pc}
\end{figure}

The angular momentum of the neutron star $L_{\rm{NS}}$ can be converted into a rotation frequency $f_{\rm{NS}}$ using the relation:
\begin{equation}
 \label{eq:fns}
 f_{\rm NS} = \frac{L_{\rm{NS}}}{2\pi I_{\rm NS}},
\end{equation}
where $I_{\rm{NS}} \sim 10^{45}\, \rm{g\ cm^{2}}$ represents the moment of inertia of the neutron star.
The rotation frequency of the core is defined in a similar way by considering 
the frequency a neutron star would have assuming a spherically-symmetric collapse of the iron core: $f_{\rm CORE} = L_{\rm{CORE}}/\left(2\pi I_{\rm NS}\right)$.
Fig. \ref{fig:Pns_Pc} shows how a spiral mode influences the relationship between the core rotation frequency and the neutron star one.
On the one hand, the results confirm that the neutron star spin period is almost not affected in cases with a low radii ratio or a high rotation rate (around the black dashed line).
On the other hand, when the radii ratio is sufficiently high and the core rotation frequency is below 100 Hz, the spiral mode is able to significantly impact the neutron star periods.
The frequencies obtained are consistent with pulsar frequencies at birth deduced from observations: $f_{\rm NS} \lesssim 100\, \rm{Hz}$. 
Both neutron star spinning-down (below the black dashed line) and neutron star spinning-up (above this line)
are possible. 

Considering a high enough radii ratio, e.g. $R\geq 3$, the influence of a spiral mode on the pulsar spin can be described as follows.
For a non-rotating progenitor, SASI is able to redistribute angular momentum and spin up a neutron star to rotation frequencies of several tens Hz 
(horizontal dashed lines in Fig. \ref{fig:Pns_Pc}).
The direction of rotation of the SASI spiral mode is stochastic, so is the neutron star one (see Paper I).
As long as the core rotation rate is negligible, the angular momentum redistribution proceeds very similarly to the non-rotating case and the initial
pulsar period is close to the one obtained without progenitor rotation (isolated points in Fig. \ref{fig:Pns_Pc}). 
In these cases, the direction of rotation of the pulsar is still stochastic since the rotation rate is not high enough 
to significantly affect the growth rates of the spiral modes \citep{yamasaki08} and
to determine the direction of rotation of the spiral.
In this range of rotation rates, the neutron star is thus not necessarily counter-rotating.
For higher rotation rates, as intuitively expected, we observe that the core angular momentum required to balance the quantity redistributed by SASI is similar
to the amount of angular momentum stored in a SASI spiral mode of a non-rotating case (see Fig. 8 of Paper I and \citet{guilet14} for an estimate). 
In such a case, the neutron star could be spun down to no rotation.
Finally, as the rotation rate increases from this specific case, the spin down driven by the spiral mode is less and less significant and reaches an asymptotic 
efficiency in cases dominated by a corotation instability (Fig. \ref{fig:Lz_TW}).

\section{Discussion}
\label{sec:discussion}

\begin{figure}
\centering
	\includegraphics[width=\columnwidth]{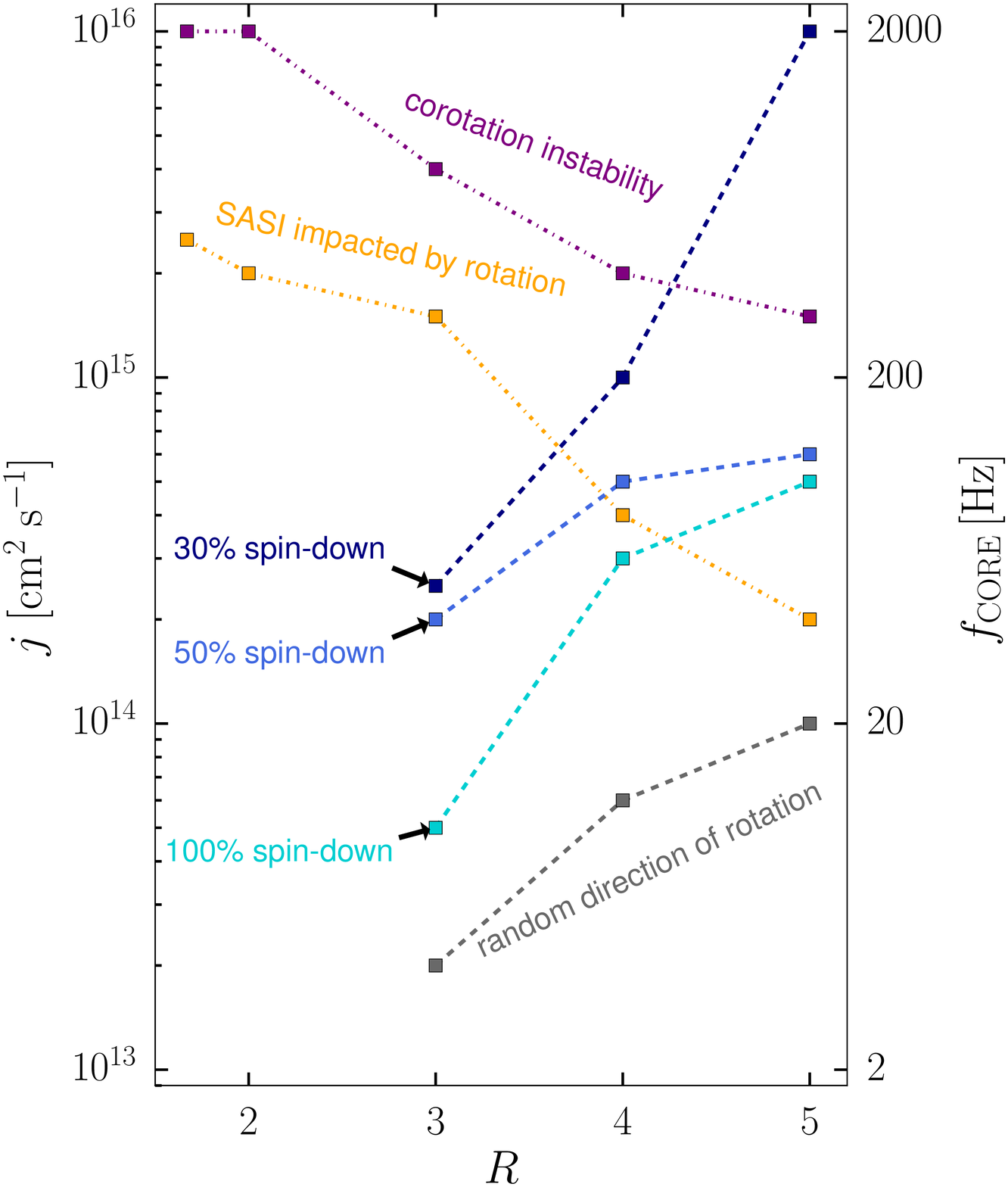}
    \caption{Different regimes in the parameter space $\left(R,\, j\right)$ regarding the influence of rotation on the shock dynamics and the neutron star spin.
    The spiral mode of SASI is affected by rotation in cases above the orange dot-dashed line. The amplitude of the spiral mode is either increased or decreased depending on the value of $R$.
    In cases above the purple dot-dashed line, a corotation radius exceeds the PNS one and a corotation instability overlaps with SASI. Below the grey dashed line, the direction of rotation of the spiral
    mode is not determined by rotation. In this region, neutron stars may either have the same direction of rotation as the progenitor or the opposite one.
    The three blue dashed lines indicate a neutron star spin-down of 30\%, 50\% and 100\% compared to a spherically-symmetric collapse. The latter value corresponds to a non-rotating neutron star.
    }
    \label{fig:regimes}
\end{figure}

The systematic exploration of the parameter space enables us to quantify together the effects of stellar rotation on the shock wave dynamics and the natal neutron star spin (Fig. \ref{fig:regimes}).
The results confirm that previous studies dedicated to the neutron star spin-up born from non-rotating progenitors \citep{fernandez10,guilet14,kazeroni16} can naturally be extended to 
slowly rotating progenitors, i.e. $j \lesssim 10^{14}\, \cms$. Below the grey dashed line in Fig. \ref{fig:regimes}, the direction of rotation of the spiral mode is not determined by the progenitor rotation.
This is compatible with the characteristics of the mechanism forming a SASI spiral mode in non-linear regime and investigated in Paper I.

If the radii ratio is such that $R<3$, the angular momentum redistribution by a spiral mode is negligible and the pulsar spin results from the conservation of the angular momentum of the iron core.
The shock radius only increases compared to the non-rotating case if a corotation instability develops. Such situations correspond to very fast rotation rates, i.e. $j \geq 10^{16}\, \cms$. 
For the rest of this section we restrict ourselves to large radii ratios $R\geq3$. In these cases, rotational effects may already become significant for a lower specific angular momentum.

As the radii ratio increases, this diminishes the minimum thresholds for rotation to boost the spiral of SASI (above the orange dot-dashed line) and to trigger a corotation instability (above the purple dot-dashed line).
These trends can be explained by an increase of the saturation amplitude and a lowering of the oscillation frequency with $R$. Besides, the maximum rotation rate to obtain a given spin-down efficiency
increases with $R$ (dashed blue curves). The intersection between the region of the parameter space in which SASI is impacted by rotation and the one in which a significant spin-down operates, e.g. 50\%,
is relatively narrow (Fig. \ref{fig:regimes}). 
Indeed, in the range of rotation rates where SASI is the only instability the dynamics is modestly impacted by stellar rotation.
For example, a case with $R \sim 4-5$ and $j \sim 5 \times 10^{14}\, \cms$
leads to a 10-15\% increase of the shock radius and a maximum spin-down of 50\%.

Fig. \ref{fig:Pns_Pc} indicates that no progenitor with a specific angular momentum higher than $j \sim 10^{15}\, \cms$ could give birth to a neutron star with a rotation frequency lower than 100 Hz. 
This suggests that if fast rotating progenitors may ease the shock revival due to a corotation instability, the resulting natal neutron star spin period may be too short to match
the distribution inferred from observations (Fig. \ref{fig:regimes}). A corotation instability does not seem to be compatible with a minimal spin period of 10 ms, at least when the magnetic field is neglected.

\section{Conclusion}
\label{sec:conclu}

The interplay of SASI and stellar rotation has been investigated by conducting a set of parametric numerical simulations of an idealised model in cylindrical geometry.
The formalism considered in this study enabled us to isolate and study the development of hydrodynamical instabilities in rotating collapsing stellar cores.
Our approach is an essential complement to state-of-the-art simulations of core-collapse supernovae where a given physical effect can not easily be disentangled
from the complexity of the modelling.
A comparison between our simulations of the linear phase of SASI and the perturbative analysis by \citet{yamasaki08} showed a good agreement, with discrepancies smaller than 15\%.
In the non linear regime, the influence of rotation on the shock wave dynamics as well as the angular momentum redistribution by one-armed instabilities have been addressed in a quantitative manner.
Our main results can be summarised as follows:
\newline

1. In the SASI-dominated regime, the saturation amplitude and the shock expansion do not depend monotonously on the rotation rate. 
These quantities either increase or decrease with increasing specific angular momentum, depending on the shock to the neutron star radii ratio (Fig. \ref{fig:shock_SASI}).
The critical radii ratio above which rotation boosts the spiral mode is similar to the threshold above which a spiral mode emerges non-linearly in non-rotating progenitors \citep{kazeroni16}.
This equality may be related to the symmetry breaking mechanism between counter-rotating spiral modes investigated in our previous study.
\newline

2. For fast enough rotation rates, $j \gtrsim 10^{15} - 10^{16}\, \cms $, a corotation radius emerges above the proto-neutron star surface (Fig. \ref{fig:regimes}). 
As a consequence, a corotation instability overlaps with SASI (Fig. \ref{fig:spiral})
and the spiral mode is greatly amplified (Fig. \ref{fig:shock_TW}) compared to cases where SASI is the only instability.
In the latter regime, the shock radius increases by $10-15\%$ compared to SASI without initial rotation while it can be amplified by $50-200\%$ in the presence of a corotation instability (Table \ref{tab:tab1}).
\newline

3. The angular momentum redistribution induced by spiral modes in rotating collapsing iron cores has been examined.
We performed a wide coverage of the parameter space to evaluate how much the natal pulsar spin period deviates from a simple estimation based on the conservation of angular momentum.
For a high enough radii ratio, i.e. $R\gtrsim 3$, SASI has the potential to significantly affect the pulsar spin if the rotation rate corresponds to 
a period of at least 10 ms inferred from a spherically-symmetric collapse. Below such a rotation rate, the rotation frequency is mostly set by SASI (Fig. \ref{fig:Pns_Pc})
and seems to be in agreement with observational data \citep{faucher06,popov12}.
In that region of the parameter space, SASI may either spin up the neutron star or spin it down, leading to a counter-rotating body. 
For the slowest rotation rates considered, i.e. $j \lesssim 10^{14}\,\cms$, the direction of rotation of the spiral mode and thus the one of the neutron star is independent of the stellar rotation.
\newline

4. The spin-down process is much less efficient in the range of rotation rates related to the corotation instability (Fig. \ref{fig:regimes}). The magnitude of the spin-down increases
with the radii ratio but does not exceed $\sim 30\%$ for $R=5$ (Fig. \ref{fig:Lz_TW}). 
Our results based on purely hydrodynamical simulations suggest that spiral modes of SASI and the corotation instability may not be able to reconcile fast rotating iron cores
with the distribution of pulsar spins at birth estimated from observations.
\newline

Our study is based on several simplifications made to deepen our physical understanding of hydrodynamical instabilities. In order to transpose our findings to more comprehensive models,
we shall first discuss the importance of the neglected ingredients. The numerical domain was restricted to the equatorial plane of a massive star in cylindrical
geometry to allow spiral modes to emerge in 2D and to perform a wider coverage of the parameter space.
Some artefacts of this simplified setup cannot be excluded and would deserve a numerical confirmation in 3D.
It has been demonstrated with various degrees of sophistication that SASI-dominated models do exist, both in 2D axisymmetric cases \citep{mueller12} and in 3D spherical simulations \citep{hanke13, melson15b}.
A strong SASI spiral mode activity can precede the shock revival as observed in a simulation of a $20\,\Msol$ progenitor \citep{melson15b}.
We expect our findings to depend weakly on the geometry considered. 
Without rotation, the minimal radii ratio $R$ for a spiral mode to prevail over sloshing modes is expected to be slightly lower in 3D spherical geometry than in 2D cylindrical one \citep{kazeroni16}.
This is a consequence of a shorter advection time and a higher growth in spherical geometry. For a given rotation rate, the growth rate of the most unstable mode is slightly higher in 3D spherical geometry \citep{blondin17}.
Spiral modes may thus be more easily destabilised in spherical geometry. Among the other simplifications of our model, the absence of neutrino heating is of particular importance. 
As the post-shock region expands, neutrino-driven convection is favoured against SASI by the long advection time which decreases the SASI growth rates and increases the $\chi$ parameter \citep{foglizzo06}.
This suggests that there exists a maximal radii ratio $R$ above which the dynamical evolution might be dominated by neutrino-driven convection. That would set an upper limit on the efficiency
of the spin-down process driven by a spiral mode at large rotation rates (Fig. \ref{fig:Lz_TW}).
\citet{iwakami14b} showed that for fast enough rotation rates a SASI-dominated model may transition to a convection-dominated model due to the expansion
of the gain region induced by rotation. These points may help to delimit the parameter space $(R,\,j)$ in which our approach remains valid when neutrino heating is included.

Our results are intended to serve as guidelines for ab initio simulations of core-collapse supernovae which include stellar rotation.
Covering a wide parameter space, the present study can be helpful to anticipate the impact of rotation on complex simulations both regarding the dynamics of SASI and the induced 
angular momentum redistribution (Fig. \ref{fig:regimes}). To this end, it may be fruitful to better constrain the radii ratio which is time and progenitor dependant.
A single ab initio simulation may not be enough to fully capture the impact of rotation on core-collapse supernovae.
Depending on the radii ratio, rotation may either increase or decrease the amplitude of the SASI spiral mode. Moreover, several realisations of a given simulation may be needed to 
account for the stochastic nature of the non-linear regime of SASI (Fig. \ref{fig:shock_SASI}).

Spiral modes do not seem to constitute a relevant spin-down mechanism for the fastest rotation rates considered in this study.
These cases, which could in principle give birth to millisecond pulsars, undergo a maximum spin-down of 30\%. That effect is insufficient to match the observed pulsar periods of several
tens to hundreds milliseconds. \citet{ott06} already pointed out the lack of known spin-down mechanisms robust enough to spin down a millisecond pulsar to a period of
at least ten milliseconds. The corotation instability does not seem to stand as a promising candidate and 
rapid rotation might not constitute a key player in the mechanism of the majority of core-collapse supernovae when the magnetic field is neglected.

Despite the simplicity of our setup, the simulations bring to light the diversity of hydrodynamical processes at play.
The overlap between SASI and the corotation instability, already noticed by \citet{kuroda14}, deserves further investigations to clearly distinguish between these instabilities at least in their 
linear regimes. Moreover, the picture may evolve when magnetic field with fast rotation is considered. 
In addition to the potential development of the magnetorotational instability \citep{akiyama03, obergaulinger09}, the dynamics of the corotation instability may be altered by the magnetic field 
\citep{fujisawa15}. The present work will be extended to study whether the magnetic field could affect the angular momentum 
redistribution process and the natal pulsar spin.

\section*{Acknowledgements}
We thank Marc Joos, S\'ebastien Fromang and Matthias Gonz\'alez for their help with the code.
We acknowledge insightful discussions with Bernhard M\"uller, Thomas Janka and Luc Dessart.
This work was granted access to the HPC resources of TGCC/CINES under the allocations x2015047094 and t2016047094 made by GENCI (Grand \'Equipement National de Calcul Intensif).
This work is part of ANR funded project SN2NS ANR-10-BLAN-0503. JG acknowledges support from the Max-Planck-Princeton Center for Plasma Physics.




\bibliographystyle{mnras}
\bibliography{rotation_SASI} 



%
%


\bsp	
\label{lastpage}
\end{document}